\documentclass[useAMS,usenatbib]{mn2e}
\usepackage{bethmacros}
\usepackage{graphicx}
\usepackage{amssymb}

\def\spose#1{\hbox to 0pt{#1\hss}}
\def\lta{\mathrel{\spose{\lower 3pt\hbox{$\mathchar"218$}}
     \raise 2.0pt\hbox{$\mathchar"13C$}}}
\def\gta{\mathrel{\spose{\lower 3pt\hbox{$\mathchar"218$}}
     \raise 2.0pt\hbox{$\mathchar"13E$}}}

\title[MaGICC Parameter Study]{Making Galaxies in a Cosmological Context:  The Need for Early Stellar Feedback}

\author[Stinson et al.]{G.\,S. Stinson$^{1}$\thanks{Email: stinson `at' mpia.de}, C. Brook $^2$, A. V. Macci\`o$^1$,  
J. Wadsley$^{3}$, T. R. Quinn$^{4}$, 
\newauthor{H. M. P. Couchman$^3$}
\vspace*{6pt}\\
$^{1}$Max-Planck-Institut f\"ur Astronomie, K\"onigstuhl 17, 69117, Heidelberg, Germany\\
$^{2}$Departamento de F\'{i}s\'{i}ca Te\'{o}rica, Universidad Aut\'{o}noma de Madrid, E-28049 Cantoblanco, Madrid, Spain\\
$^{3}$Department of Physics and Astronomy, McMaster University, Hamilton, Ontario, L8S 4M1, Canada\\
$^{4}$Astronomy Department, University of Washington, Box 351580, Seattle, WA, 98195-1580}

\begin{document}
\maketitle
\label{firstpage}

\begin{abstract}
We introduce the Making Galaxies in a Cosmological Context (MaGICC) program of smoothed particle hydrodynamics (SPH) simulations.  We describe a parameter study of galaxy formation simulations of an L$^\star$ galaxy that uses early stellar feedback combined with supernova feedback to match the stellar mass--halo mass relationship.  While supernova feedback alone can reduce star formation enough to match the stellar mass--halo mass relationship, the galaxy forms too many stars before $z=2$ to match the evolution seen using abundance matching.  Our early stellar feedback is purely thermal and
thus operates like a UV ionization source as well as providing some additional pressure from the radiation of massive, young stars.  The early feedback heats gas to $>10^6$ K before cooling to $10^4$ K.  The pressure from this hot gas creates a more extended disk and prevents more star formation prior to $z=1$ than supernovae feedback alone.   The resulting disk galaxy has a flat rotation curve, an exponential surface brightness profile, and matches a wide range of disk scaling relationships.  The disk forms from the inside-out with an increasing exponential scale length as the galaxy evolves.  Overall, early stellar feedback helps to simulate galaxies that match observational results at low and high redshifts.
\end{abstract}


\section{Introduction}
As matter collapses in the early Universe, gas initially heats as kinetic energy is converted into thermal \citep{Rees1977, Birnboim2003}.  In massive halos, the gas reaches temperatures where its cooling time becomes longer than the Hubble time \citep{Rees1977}, which results in a theoretical maximum galaxy mass.  However, in the centers of galaxies, enough gas accumulates to increase the density to the point that the gas efficiently radiates and cools.  Such cooling is unstable since the radiation leaves through the optically thin surrounding hot gas.  This process leads to overcooling since as the hot gas cools, it stops providing pressure support for the surrounding gas, more dense gas is able to accumulate and cool even more efficiently \citep{White1978}.  This problem is known as the ``overcooling catastrophe'' \citep[e.g.][]{Balogh2001}.

The ``overcooling catastrophe'' has long plagued simulations of disk galaxy formation.  Fully cosmological numerical galaxy simulations consistently contain a massive central concentration of stars \citep{Navarro1991, Governato2004, Stinson2010, Scannapieco2012}.  This is evident in the central peak in rotation curves in simulated galaxies as well as in their surface brightness profiles.  

Further evidence for simulations forming too many stars comes from recent studies matching observations of galaxy stellar masses with their dark matter halo masses.  One method is the abundance matching technique.  To lay the groundwork for this technique, \citet{Conroy2006} rank ordered halos by total halo mass from collisionless simulations and then did a rank ordering of galaxies based on their stellar mass from the Sloan Digital Sky Survey.  The halos and galaxies were then divided in several mass bins.  \citet{Conroy2006} found that the correlation functions between halos in certain mass bins were well matched by the correlation functions of galaxies in corresponding stellar mass bins.  Several groups used this abundance matching technique to compare total halo masses with galaxy stellar masses \citep{Conroy2009,Moster2010,Guo2010,Behroozi2010}.  Direct comparisons of stellar mass and total halo mass have also been done based on satellite dynamics \citep{Klypin2009, More2009, More2011} and weak lensing \citep{Mandelbaum2009, Schulz2010}.  For a comparison of all these techniques, please see Figure 11 of \citet{Behroozi2010}.  Each method shows a good level of correspondence.  \citet{Guo2010} and \citet{Sawala2011} showed that nearly all the simulations of galaxy formation have formed many more stars than abundance matching predicts.  

Even without overcooling, \citet{vandenBosch2002} showed that the amount of low angular momentum material in collapsed collisionless halos exceeds the amount of low angular momentum material observed in disk galaxies.  Consequently, this low angular momentum material needs to be removed from the center of the system.  

Stellar feedback is the favored way of reducing  star formation and launching outflows \citep{Scannapieco2008, Schaye2010}.  \citet{Dekel1986} showed that supernova feedback can eject gas from galaxies with virial velocities up to 100 km s$^{-1}$.  Semi-analytic models have found that significant amounts of stellar feedback are required to match the low mass end of the luminosity function \citep{Somerville1999, Benson2003, Bower2006, Bower2008, Bower2012}. \citet{Dutton2009} showed that stellar feedback can remove low angular momentum material.  

In hydrodynamical simulations, two methods are commonly used to model stellar feedback.  One is kinetic feedback that adds velocity kicks to gas particles to remove them from the inner regions of galaxy disks \citep{SH03, Oppenheimer2006, DallaVecchia2008}.  The other is thermal feedback in which stars simply heat gas particles and allow the adiabatic work of the particles to push other gas out of the way \citep{Gerritsen1997, Thacker2000, Kawata2003, Stinson2006}.  

Since stars form in dense regions, the cooling times of the surrounding gas are short, and without help, the gas will quickly radiate away all the supernova energy \citep{Katz1992}.  In real galaxies, the amount of gas necessary to exert a dynamical influence on the ISM small.  In simulations, such small amounts of gas are difficult to model, so a common technique has been to turn off cooling for a limited amount of time \citep{Gerritsen1997, Thacker2000, Brook2004, Stinson2006}.  

It is not clear that kinetic and thermal feedback have significantly different effects than one another.  \citet{Durier2012} showed that kinetic feedback has the same effect as thermal feedback when the hydrodynamics is left turned on.  However, in many implementations of kinetic feedback, the hydrodynamic processes are disabled for a set period of time to maintain numerical convergence \citep{SH03, Oppenheimer2006}.  

It is also possible that thermal feedback does not need to rely on disabling cooling.  \citet{DallaVecchia2012} recently showed that thermal feedback can have a significant dynamical effect if the total energy of all the supernovae explosions is injected into one particle at one time.  This large, one time energy deposition raises the particle's temperature high enough that its cooling time lengthens enough for the feedback to have a dynamical effect.

For the first time in simulations, \citet{Governato2010} showed that thermal stellar feedback can remove low angular momentum material from dwarf galaxies, creating galaxies with slowly rising rotation curves.  \citet{Brook2011} showed that the ejected low angular momentum material becomes part of a galactic fountain and can be reaccreted onto the disk with higher angular momentum than it left.  
\citet{Sawala2011} showed that even using a high density threshold for star formation that made the stellar feedback more efficient, these simulations produced more stars than is predicted by the stellar mass--halo mass relationship.  

Three other recent simulations have shown success at forming realistic $L^\star$ disk galaxies.  \citet{Guedes2011} used high resolution and a high amount of supernova feedback to produce a realistic disk galaxy.  \citet{Agertz2011} suggested that using a low star forming efficiency is helpful for creating extended disks.  They also use models that increase the amount of supernova energy deposited above the canonical $10^{51}$ erg per supernova.  These simulations do the best job flattening out the central region of their rotation curve.  \citet{McCarthy2012} also form galaxies with flat rotation curves in a large, low resolution cosmological volume.  They employ a kinetic feedback scheme that uses less than the canonical $10^{51}$ erg supernova energy, though see \citet{Kay2002} for a demonstration of how kinetic feedback provides a larger effect than thermal feedback.  The galaxies each contain a few thousand particles and employ a gravitational softening of $\sim1$ kpc, which makes an examination of the detailed structure of the galaxies challenging.

These simulations show that it is possible that more feedback than the canonical $10^{51}$ erg is required.  Instead of simply increasing the amount of energy released by supernovae feedback, we note that \citet{Murray2010} showed that there can be significant feedback effects from stars before they explode as supernovae.  \citet{Hopkins2011} incorporated a kinetic radiation pressure feedback into simulations of isolated disk galaxies and found that the feedback could strongly regulate star formation.  While we do not have the resolution in cosmological simulations to implement a similar kinetic scheme, we implement a scheme based on thermal pressure to provide feedback during the time between when stars are formed and the first SN star exploding.

Using this feedback prescription, \citet{Maccio2012} showed that the feedback removes low angular momentum dark matter in galaxies up to nearly L$^\star$, producing cored dark matter density profiles.  
\citet{Stinson2012} showed that the metal rich outflows created by this feedback match observations of OVI in the circum-galactic medium of star forming galaxies.  \citeauthor{Brook2012a} (2012b) also showed that
a sample of lower mass galaxies form disks that follow a wide range of disk scaling relationships.

Here, we present a detailed study of how varying the key free parameters in our simulations affect the morphology and evolution of galaxies.  In \S \ref{sec:sims}, we describe the galaxy we model and the physics used in the simulation.  In \S \ref{sec:results}, we show how varying parameters affects the mass of stars formed, the morphology of the galaxy, and how the galaxy evolved.  

\section{Simulations}
\label{sec:sims}
We use g1536, a cosmological zoom simulations drawn from the McMaster Unbiased Galaxy Simulations (MUGS), in our parameter study.  See \citet{Stinson2010} for a complete description of the creation of the initial conditions.  
g1536 has a total mass of $7\times10^{11}$ M$_\odot$, a spin parameter of 0.017, and a last major merger at $z=2.9$.  Using the physics employed in the original MUGS simulations, g1536 also formed an exponential disk with an classical bulge with a central surface brightness of $\mu_i=14$, a total face-on magnitude of M$_r$=-21.7 and $g-r$ colour of 0.62.  

Using the smoothed particle hydrodynamics (SPH) code \textsc{gasoline} \citep{Wadsley2004}, we varied the stellar feedback to reduce star formation to the amount prescribed in \citet{Moster2012}.  We first increased the energy input from supernovae to $10^{51}$ erg and changed the initial stellar mass function (IMF) from what was described in \citet{Kroupa1993} to an IMF that contains more massive stars as described in \citet{Chabrier2003}.  Neither of these changes sufficiently limited star formation, so we turned to a source of energy that is often ignored in cosmological simulations, the luminosity emitted by massive stars before they explode as supernovae.  \citet{Hopkins2011} has shown that the inclusion of radiation pressure in high resolution simulations can reduce star formation.  The supernovae explosions in our previous simulations did not happen until 4 Myr after the formation of a stellar population.  However, molecular clouds are disrupted much sooner than this, so the inclusion of this early stellar feedback is necessary to reduce star formation before supernovae start exploding.

\subsection{Gas Cooling}
In addition to energy input, the gas cooling has a large effect on how many stars form.  The cooling used in this paper is described in detail in \citet{Shen2010}.  It was calculated using \textsc{cloudy} (version 07.02; \citet{Ferland1998}) including photoionization and heating from the \citet{Haardt2005} UV background, Compton cooling, and hydrogen, helium and metal cooling from 10 to $10^9$ K.  \citet{Shen2010} showed that $T>10^4$ K metal cooling including the photoionization effects of the UV background reduces cooling by an order of magnitude from purely collisional cooling rates \citet{Sutherland1993}.  

In the dense, interstellar medium gas, we do not impose any shielding from the extragalactic UV field.  While the cosmic UV background used is
lower than the typical UV field within galaxies, it does provide heating
similar to that expected in the ISM.

When metal cooling is included below $10^4$ K, it leads to the well-known two-phase instability of the ISM \citep{Field1965, Field1969}, which produces densities over $100$ cm$^{-3}$ and temperatures of 100 K in these simulations.  These values correspond to a Jeans mass of $<$6000 M$_\odot$ and a Jeans length of $<$8 pc.  Each of these values are far below both the gas particle mass, $2\times10^5$ M$_\odot$, or 310 pc softening length.  In reality, this phase would become even denser, self-shielded, and ultimately molecular, but we cannot resolve this process and must therefore estimate the star formation rate from the resolved dense gas.  To avoid issues with Jeans fragmentation, a minimum pressure is established in the gas as described in \citet{Robertson2008}.  A maximum density is also set to 400 cm$^{-3}$ using a minimum smoothing length $\frac{1}{4}$ of the 310 pc gravitational softening.

\renewcommand{\thefootnote}{\alph{footnote}}
\begin{table*}
\label{tab:data}
\begin{minipage}{180mm}
\begin{center}
\caption{Simulation data}
\begin{tabular}{llllllll}
\hline
Name & color & c$\star$\footnotemark[1] & $\epsilon_{esf}$\footnotemark[2] & 
$C_{td}$\footnotemark[3] & IMF\footnotemark[4] & M$_\star$ [M$_\odot$]\footnotemark[5] & 
$M_V$\footnotemark[6] 
 \\
\hline
Fiducial & red & 0.1 & 0.1 & 0.05 & C & $2.3\times10^{10}$ & -20.6 \\
MUGS & black & 0.05 & 0 & 0.05\footnotemark[7] & K & $8.3\times10^{10}$ & -21.4 \\
Low Diffusion & green & 0.1 & 0.1 & 0.01\footnotemark[7] & C & $3.0\times10^{10}$ & -21.1 \\
High Diffusion & yellow & 0.1 & 0.175 & 0.05\footnotemark[7] & C & $2.5\times10^{10}$ & -21.4\\
No ESF & magenta & 0.1 & 0 & 0.05 & C & $3.4\times10^{10}$ & -19.5 \\
120\% SN energy & cyan & 0.1 & 0 & 0.05 & C & $1.8\times10^{10}$ & -20.0 \\
High ESF & blue & 0.1 & 0.125 & 0.05 & C & $1.1\times10^{10}$ & -19.9 \\
\hline
\end{tabular}
\footnotetext[1]{Star forming efficiency}
\footnotetext[2]{Early Stellar Feedback Efficiency }
\footnotetext[3]{Thermal Diffusion coefficient  }
\footnotetext[4]{Initial Mass Function: C=\citet{Chabrier2003}; K=\citet{Kroupa1993}}
\footnotetext[5]{M$_\star$ is the total stellar mass}
\footnotetext[6]{V-band magnitude}
\footnotetext[7]{All particles diffused thermal energy including those with cooling shut off}
\end{center}
\end{minipage}
\end{table*}

\subsection{Star Formation and Feedback}
\label{sec:feedback}
The simulations use a common recipe for star formation described in \citet{Stinson2006} that we summarize here.  Stars form from cool ($T < 15,000$ K), dense gas.  The metal cooling readily produces dense gas, so the star formation density threshold is set to the maximum density at which gravitational instabilities can be resolved, $\frac{32 M_{gas}}{\epsilon^3}$($n_{th} > 9.3$ cm$^{-3}$), where $M_{gas}=2.2\times10^5$ M$_\odot$ and $\epsilon$ is the gravitational softening (310 pc).  Such gas is converted to stars according to the equation
\begin{equation}
\frac{\Delta M_\star}{\Delta t} = c_\star \frac{M_{gas}}{t_{dyn}}
\end{equation}
Here, $\Delta M_\star$ is the mass of the star particle formed, $\Delta t$ is the timestep between star formation events, $8\times10^5$ yr in these simulations, and $t_{dyn}$ is the gas particle's dynamical time.  $c_\star$ is the efficiency of star formation, in other words, the fraction of gas that will be converted into stars during $t_{dyn}$. 

Stars feed both energy and metals back into the interstellar medium gas surrounding the region where they formed.  Supernova feedback is implemented using the blastwave formalism described in \citet{Stinson2006}.  In this, Type II supernovae are assumed to explode due to the core collapse at the end of lifetime of stars greater than 8 M$_\odot$.  Stellar lifetimes are based on the Padua lifetimes \citep{Raiteri1996}.  Since the gas receiving the energy is dense, it would quickly be radiated away due to its efficient cooling.  For this reason, cooling is disabled for particles inside the blast region 
\begin{equation}
r_{CSO} = 10^{1.74}E_{\rm 51}^{0.32}n_0^{-0.16}\tilde{P}_{\rm 04}^{-0.20} {\rm pc}
\end{equation}
and for the length of time 
\begin{equation}
t_{CSO} = 10^{6.85}E_{\rm 51}^{0.32}n_0^{0.34}\tilde{P}_{\rm 04}^{-0.70} {\rm yr} 
\end{equation}
given in \citet{McKee1977}.  

\subsection{Early Stellar Feedback}
In an effort to provide a more physical model for stellar feedback, radiation energy from massive stars is now considered.  Heating is introduced immediately after stars form based on how much star light is radiated.  The early energy input from high mass stars has been entirely ignored in most previous cosmological simulations.  In such cases, 4 Myr pass after stars form before the first supernova explodes \citep[some simulations use even longer delays, see][]{Slyz2005, DallaVecchia2012}, during which time stars can continue forming without any affect of stellar feedback.  

\citet{Hopkins2011} describes how such energy could be considered using a kinetic scheme where the radiation pressure drives winds out of massive star clusters.  Our simulations do not have the resolution to use kinetic feedback because it requires the choice of a direction.  A kinetic implementation of feedback requires many particles to represent a molecular cloud so that it can be torn apart.  In our simulations, only one or a few particles represent a molecular cloud, so we utilize thermal pressure, which is isotropic.  Using thermal feedback provides pressure support and increases gas temperatures above the star formation threshold to decrease star formation.  We do not disable cooling during these early times since massive young stars radiate a large amount of energy.  Thus, after heating the gas to $T>10^6$ K, the gas rapidly cools to $10^4$ K, which creates a lower density medium than if the gas were allowed to continue cooling until supernovae exploded.

To model the luminosity of stars, a simple fit of the mass-luminosity relationship observed in binary star systems by \citet{Torres2010} is used:
\begin{equation}
\,\frac{L}{L_\odot} =  \left\{
\begin{array}{ll}
\left(\frac{M}{M_\odot}\right)^{4} &, M < 10 M_\odot \, \\
100\left(\frac{M}{M_\odot}\right)^{2} &, M > 10 M_\odot 
\end{array}\right.
\end{equation}
Typically, this relationship leads to $2\times10^{50}$ erg of energy being released from the high mass stars per M$_\odot$ of the entire stellar population over the 4 Myr between the star's formation and the commencing of SNII (SNII inject $\sim10^{49}$ erg per $M_\odot$).  

\citet{Leitherer1999} found that approximately 10\% of the total stellar flux is emitted in the UV, so we use with an early stellar feedback efficiency, $\epsilon_{esf}$=10\%.  \citep{Freyer2006} found that stellar photons do not couple efficiently with the surrounding ISM.  Correspondingly, radiative cooling is \emph{not} turned off for this form of energy input. The dense gas thus cools rapidly to $10^4$ K in the ISM, which mimics the formation an HII region.  Though the dynamical effect is minimal, early stellar feedback effectively halts star formation in the region immediately surrounding a recently formed star.  Using $\epsilon_{esf}$=10\% limits star formation to the amount prescribed by the stellar mass--halo mass relationship.  

\subsection{Diffusion}
\citet{Agertz2007} highlighted how long cold gas clumps can survive in a surrounding bath of hot gas in SPH relative to Eulerian grid simulations.  In an effort to decrease this mixing time, \citet{Wadsley2008} implemented diffusion into SPH based on mixing that should result in shearing flows.  The amount of mixing depends on the magnitude of the local velocity shear field multiplied by the square of a length comparable to the grid or resolution scale.  \citet{Wadsley2008} showed that  diffusion of thermal energy effectively limits the lifetime of the blobs that \citet{Agertz2007} described.  In MUGS \citep{Stinson2010}, both metals and heat diffused using this scheme with a diffusion coefficient of 0.05.  The most striking effect of the thermal diffusion was that cold accretion filaments were heated and expanded.  Cold blobs persisted in these simulations.  Results of varying the thermal diffusion coefficient are presented in \S \ref{sec:results}.  

During this study, it was determined that the heat diffusion severely reduced the efficiency of the adiabatic feedback scheme to drive outflows.  Hot particles with their cooling shut off frequently passed cooler particles with their cooling turned on.  Heat would diffuse from the hot to cold particles, reducing the temperature of the particles and thus the efficiency of the feedback.  This would happen mostly as hot particles orbited around the disk and occasionally after particles had been ejected from the disk. Since the feedback is dependent on particles not cooling, we disabled thermal diffusion in interactions between particles in which either particle had its cooling shut off.  We note that disabled cooling has the biggest effect in the dense disk.  We find less than 1\% of hot (T$>10^4$ K) particles with their cooling disabled are outside the disk at $z=0$ and such particles are never found beyond 25 kpc at any time.  

\begin{figure}
\resizebox{9cm}{!}{\includegraphics{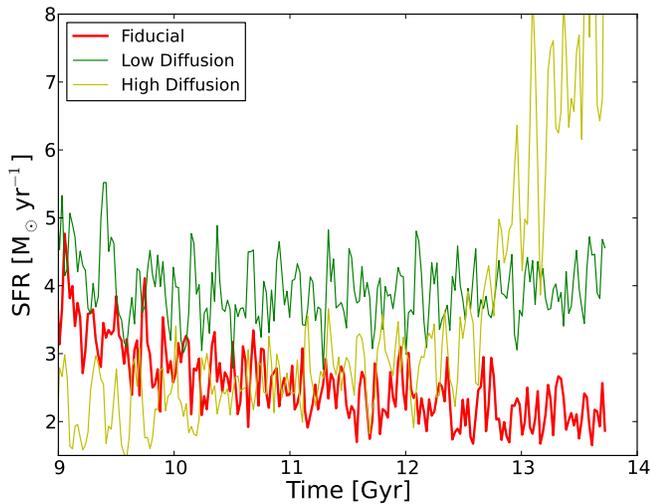}}
 \caption[Diffusion SFH]{ The star formation history of galaxies simulated with varying amounts of thermal diffusion.  Here, the simulation evolves from left to right.  This plot focuses on the last 5 Gyr of the simulations since that is when the biggest difference in star formation rate is seen.}
\label{fig:diffsfh} 
\end{figure}

Figure \ref{fig:diffsfh} shows how the star formation histories of the different implementations of thermal diffusion vary.  In the fiducial model, the star formation declines until the end of the simulation.  The scheme where all gas particles always diffused required a 75\% increase in $\epsilon_{\rm ESF}$ to limit the star formation to the stellar mass--halo mass relationship.  The shape of the star formation history for the high diffusion model is different.  It increases throughout the history of the galaxy and begins to start a starburst over the last 1 Gyr.  This shows how with too much thermal diffusion star formation acts as a positive feedback:  the more stars form, the more gas cools enabling more star formation.  The low thermal diffusion simulation (\emph{green}) demonstrates a milder increase in star formation rate at the end of the simulation, but it remains above the fiducial value.  An examination of the simulation reveals that this is due to cold gas clouds cooling out of the hot halo as described in \citet{Maller2004} and \citet{Kaufmann2009}.  In the revised treatment, the limited diffusion helps to limit the formation of cold blobs and reduces the star formation rate at late times in the simulation.

\section{Results}
\label{sec:results}
\begin{figure}
\resizebox{9cm}{!}{\includegraphics{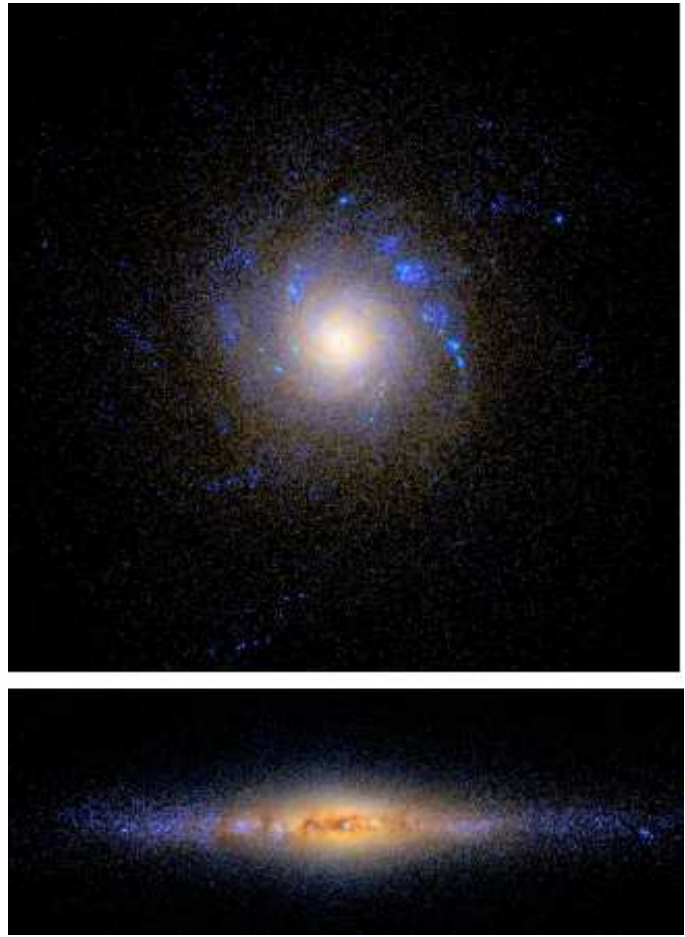}}
 \caption[SUNRISE image]{ Face-on and edge-on images of the fiducial galaxy at $z=0$.  The images are 50 kpc on a side and were created using the Monte Carlo radiative transfer code \textsc{sunrise}.  The image brightness and contrast are scaled using $asinh$ as described in \citet{Lupton2004}. }
\label{fig:sunrise} 
\end{figure}

To find an optimal star formation and feedback recipe, we attempt to match the stellar mass--halo mass relationship.  The parameters for all of our simulations are presented in Table 1.  Figure \ref{fig:sunrise} shows a mock observation of the fiducial galaxy.  The image is 50 kpc on a side and was created using the Monte Carlo radiative transfer code \textsc{sunrise} \citep{Jonsson2006}.  The image brightness and contrast are scaled using $asinh$ as described in \citet{Lupton2004} since disks have an exponential surface brightness profiles meaning the images span a wide range in surface brightness.  The image shows a thin, young stellar disk with clumpy and extended star formation, a dust lane, and a red bulge component.\footnote{Movies of the simulations can be found at http://www.mpia.de/$\sim$stinson/magicc.}

\subsection{Total Mass}
\begin{figure}
\resizebox{9cm}{!}{\includegraphics{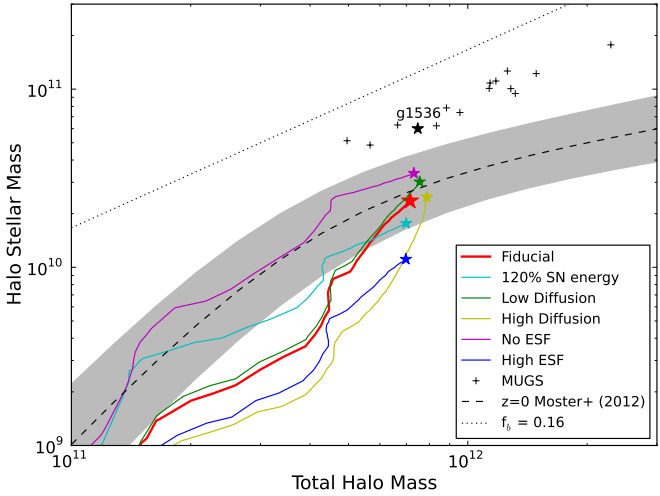}}
 \caption[Stellar Mass vs Halo Mass]{ Stellar mass plotted as a function of halo mass.  The abundance matching fit from \citet{Moster2012} and the cosmic baryon fraction lines are shown as reference.  All the MUGS simulations formed too many stars along a line where half the baryons have turned into stars and half are in the hot gas halo.  The simulations of g1536 are shown as coloured stars and lines.  The legend provides a brief description of which simulation each of these colours represents, while a more extensive description can be found in Table 1.  The fiducial simulation is coloured red and lies closest to the \citet{Moster2012} relationship.  The yellow star representing the simulation where all gas particles diffuse thermal energy also lies close to the relationship line, but it is rapidly forming stars and will quickly overshoot the relationship.  The blue point uses the same physics as the fiducial run, but increases the $\epsilon_{esf}$ by 25\% and forms far fewer stars.  The magenta line shows a simulation run without early stellar feedback.  While it ends up close to the fiducial run at $z=0$, it follows a dramatically different evolutionary path.}   
\label{fig:guo} 
\end{figure}
Figure \ref{fig:guo} shows the mass of the stellar component at $z=0$ as a function of halo mass.  For the sake of clarity we have limited the number of simulations displayed to those which demonstrate our walk through parameter space.  The coloured stars show the $z=0$ stellar and halo masses of the simulations of g1536 used in this parameter study.  Each $z=0$ point is accompanied by a line that shows the path along which it evolved through this diagram.  A couple of lines are provided for reference.  The dotted line is the cosmic baryon fraction of the total halo mass.  The dashed line is the stellar mass halo mass relationship found in \citet{Moster2012}.  The black plus signs are the old results from MUGS.  These lower feedback runs are similar to many other low feedback simulations where galaxies turn about half of a galaxy's baryons, as predicted by the cosmic baryon fraction, into stars.  

The red star represents the best fit to the stellar mass--halo mass relationship.  It includes early stellar feedback with thermal diffusion disabled for particles with their cooling disabled.  It uses a star formation efficiency, $c^\star=$0.1, an early stellar feedback efficiency, $\epsilon_{esf}=$0.1, and a thermal diffusion coefficient, $C_{td}=$0.05.  The fiducial simulation evolves along a path generally below the $z=0$ \citet{Moster2012} relationship, which is shown in \S \ref{sec:sfh} to be similar to how the observed stellar to halo mass ratio evolves.  A small change in the amount of feedback causes a significant change in the number of stars that form.  When $\epsilon_{esf}$ increases from 0.1 to 0.125 (represented in blue), the stellar mass decreases by a factor of 2.  

The magenta star and line represent the simulation in which only the supernova feedback was increased from the MUGS value of $4\times10^{50}$ erg per supernova to $10^{51}$ erg.  In this simulation, the thermal diffusion is turned off for particles with their cooling disabled.  This change significantly decreases the mass of stars formed, though the stellar mass still lies above the \citet{Moster2012} relationship.  This motivated us to add early stellar feedback.

As a check to see whether the addition of more energy or the timing of that energy addition had a stronger effect, we ran a series of simulations without the $10^{51}$ erg per supernova energy constraint.  The best result from that series is represented by the $cyan$ line and star.  It shows that a 20\% increase of the supernova energy lowered the stellar mass to the \citet{Moster2012} relationship.  Its evolution more closely follows the $z=0$ \citet{Moster2012} relationship than the simulation using $10^{51}$ erg.  Again, we refer the reader to \S \ref{sec:sfh} to see that the evolution does not follow what is described in \citet{Moster2012}.

The yellow star and line show the reduced effect of stellar feedback when thermal diffusion is active for particles with their cooling shutoff.  This requires the early stellar feedback efficiency, $\epsilon_{esf}$, to be increased to 0.175 to produce a galaxy that matches the stellar mass--halo mass relationship at $z=0$.  However, its evolution is much different from the other galaxies.  Its stellar mass is increasing sharply at $z=0$.  Figure \ref{fig:diffsfh} shows that its star formation rate is also dramatically increasing, so it will quickly overshoot the relationship.  The drastic increase in star formation rate is the result of thermal feedback diffusing away instead of contributing to launching winds.

Simply reducing the thermal diffusion for all gas particles from 0.05 to 0.01, as shown by the green star, required a corresponding reduction to $\epsilon_{esf}$ to form enough stars to match the stellar mass--halo mass relationship.  However, the reduced thermal diffusion allowed cold clouds to condense out of the hot halo at the end of the simulation, which again led to a increasing star formation rate, though not as dramatic as in the higher thermal diffusion case.

As we made our parameter study, we found a degeneracy between $c^\star$ and $\epsilon_{esf}$, so that a simulation where $c^\star$=0.05 and $\epsilon_{esf}$=0.15 produces the same results as a simulation where $c^\star$=0.1 and $\epsilon_{esf}$=0.1.  Such a degeneracy is not surprising as the early feedback is immediate and limits star formation efficiency.

\subsection{Mass Distribution}
\begin{figure}
\resizebox{9cm}{!}{\includegraphics{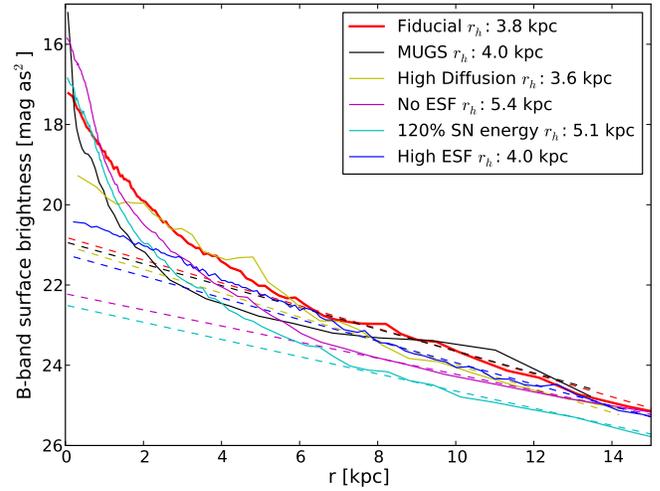}}
 \caption[Surface Brightness Profile]{ B-band surface brightness profile for g1536 simulated 
at a variety of feedback strengths.  Each surface brightness profile has an exponential component
with a central surface brightness around 21 mag arcsec$^{-2}$. All the simulations that 
contain too many stars in Figure \ref{fig:guo} have an exceptionally steep central 
profile that ends at a central surface brightness brighter than most observed galaxies.
The galaxy that uses the MaGICC feedback deviates slightly from the pure exponential line, but
does not increase nearly as quickly as the simulations with lower feedback.  }
\label{fig:sb} 
\end{figure}

Having used the stellar mass--halo mass relationship as our constraint, we now examine other properties of the galaxies.  Figure \ref{fig:sb} shows surface brightness profiles produced using different amounts of stellar feedback in g1536.  The one dimensional, face-on surface brightness profiles were constructed using the stellar population models from \citet{Girardi2010}.  Only stars in a disk with a radius of 15 kpc and height 2 kpc above and below the disk midplane are included.  The profiles represent the mean values of 100 linearly spaced azimuthal bins.  An exponential component of the disk was fit using stars with radii greater than 5 kpc.  These fits are represented by the dashed lines the same color as the surface brightness profiles.  Each exponential fit has a scale length around 4 kpc and central surface brightnesses around 21 in the B-band, slightly brighter than the \citet{Freeman1970} Law.

The fiducial simulation forms the right amount of stars and has an exponential surface brightness profile that steepens at 5 kpc to reach a central $B$ surface brightness of 17 mag arcsec$^{-2}$.  By contrast, the simulations of g1536 without early stellar feedback each have exponential surface brightness profiles with slightly longer scale lengths but more extensive bulges.  These bulges are characterized by higher Sersic index components in their inner regions that reach brighter central surface brightnesses than the simulations with early stellar feedback.  This high central surface brightness is a common symptom of excess stellar mass in the central galactic region found in many galaxy simulations \citep{Scannapieco2010, Stinson2010, Scannapieco2012}.

\begin{figure}
\resizebox{9cm}{!}{\includegraphics{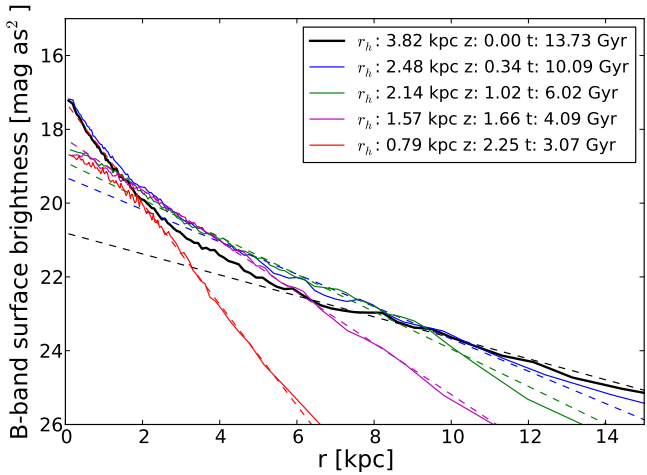}}
 \caption[Surface Brightness Evolution]{ The evolution of the B-band surface brightness profile in the fiducial simulation. }
\label{fig:sbev} 
\end{figure}
Figure \ref{fig:sbev} shows the evolution of the $B$ surface brightness profile in our fiducial model.  It shows the disk growing from the inside out.  For the bulk of its evolution, the surface brightness profile remains exponential. In the final two time steps shown, an excess develops in the center.
The scale length of the disk undergoes an evolution from 1 kpc at $z=2$ to 4 kpc at $z=0$.

\begin{figure}
\resizebox{9cm}{!}{\includegraphics{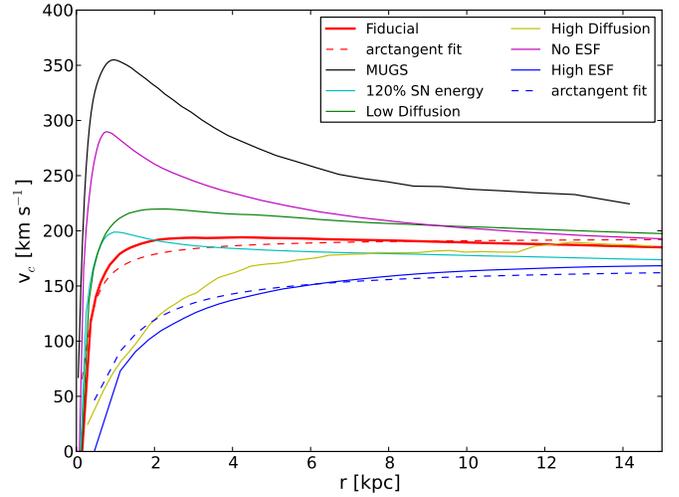}}
 \caption[Rotation Curve Comparison]{ Plot of $v_c=\sqrt{GM(r)/r}$ as a function of $r$ for the all the models of g1536 and g5664 at $z=0$ modeled with varying amounts of stellar feedback.  }
\label{fig:rotcur} 
\end{figure}
To examine the underlying mass distribution of the galaxy, Figure \ref{fig:rotcur} shows the rotation curves, the circular velocity, $v_c =  \sqrt{GM(r)/r}$, as a function of radius, for the same galaxy models.  The fiducial model has a nearly flat rotation curve.  The rotation curve for the MUGS g1536 has a large central peak.  The simulation without early stellar feedback also exhibits a high central peak in its rotation curve.  However, the simulation with 20\% more supernova energy has a much flatter rotation curve, though not as flat as the fiducial model.

The simulations with too much feedback have slowly rising rotation curves that are in conflict with the nearly flat rotation curve observed for the Milky Way and other L* galaxies \citep{Courteau1997, deBlok2008}.  We have drawn sample model fits for the fiducial and high early stellar feedback simulations following \citet{Reyes2011}.  They use $R_{TO}$ values, radius at which the profile turns over, of 0.25 and 1 kpc, respectively.  The slowly rising rotation curves show the best agreement with the arctangent models used in \citet{Reyes2011}, where typical $R_{TO}$ values are approximately the same as the disk scale length for galaxies with stellar masses similar to g1536's $2\times10^{10}$ M$_\odot$, which means that the rotation curves do not reach their peak until well beyond the disk scale length.  

Even though the rotation curves slowly rise in the plotted region, the peak of the rotation curve, 180 km s$^{-1}$ is still higher than the value of $v_{vir}$, 110 km s$^{-1}$.  In other words, the rotation curves turn over and decline outside the plotted region.

\begin{figure}
\resizebox{9cm}{!}{\includegraphics{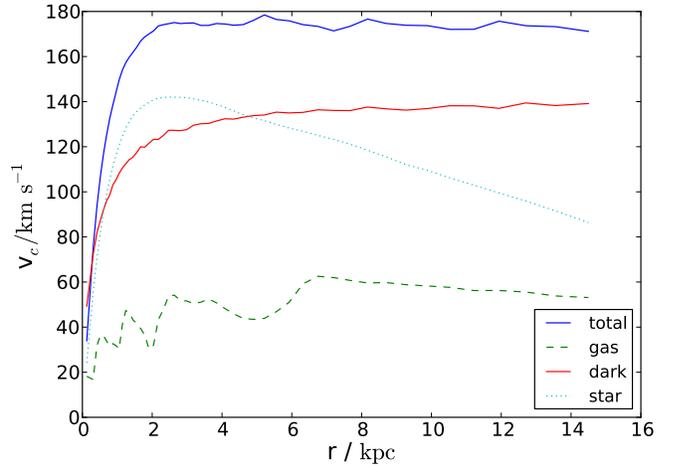}}
 \caption[Rotation Curve]{ Plot of $v_c=\sqrt{GM(r)/r}$ as a function of $r$ for the simulated galaxy at $z=0$.  Each component, dark matter (\emph{solid red}), gas (\emph{green, dotted}), and stars (\emph{light blue, dashed}), is plotted separately to show the matter distributions of each component.  }
\label{fig:rcparts} 
\end{figure}

The rotation curve for the fiducial galaxy is separated into the contributions from its constituent particle types in Figure \ref{fig:rcparts}.  Stars dominate the potential in the central regions of the galaxy.  The curves represent the mass included from just that component, so it does \emph{not} represent the speed at which those particles are orbiting.  Each component will orbit close to the total velocity.  

\begin{figure}
\resizebox{9cm}{!}{\includegraphics{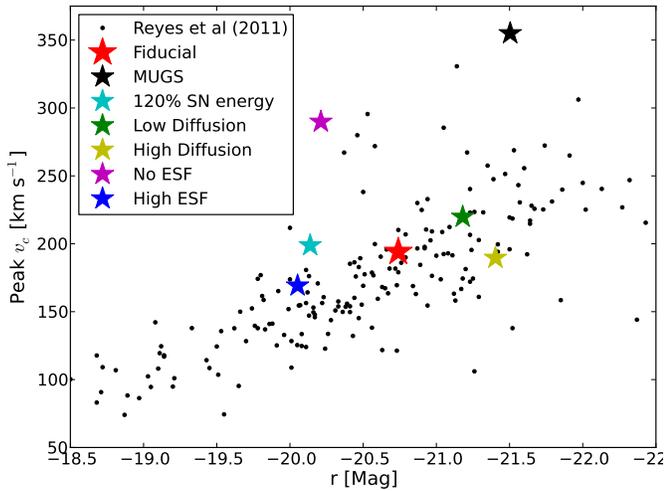}}
 \caption[Tully-Fisher Relationship]{Peak circular velocity as a function of $r$-band magnitude compared to the \citet{Reyes2011} observations.  The simulations are plotted with the standard colour scheme used throughout this paper.  The fiducial simulation ($red$) lies directly in the center of the observations.  The simulations without early stellar feedback lie significantly above the observations.}
\label{fig:tf} 
\end{figure}

The rotation velocity information enables another comparison between the amount of stars that have formed and the mass of the galaxy, the Tully-Fisher relationship \citep{Tully1977}.  The Tully-Fisher relationship compares the luminosity of a galaxy with its circular velocity.  It is possible to argue for the choice of many different radii at which to measure the circular velocity.  In \citet{Stinson2010}, the galaxies were reported as fitting the Tully-Fisher relationship because the circular velocities were taken at 2.2 $R_d$, where $R_d$ is the scale radius, which was outside the central velocity peak.  Here, we instead choose to compare our simulations with \citet{Reyes2011}, who use an arctangent model and report the peak velocities found using that fit.  We compare these with the peak velocity from our simulations.  

The fiducial simulation along with the other simulations that use early stellar feedback lie in the middle of the observed relationship.  One exception is the high diffusion simulation, which has a bright $r$-band magnitude due to its significant late star formation but still has a low peak velocity.  The simulations without early stellar feedback lie above most of the galaxies in the observed relationship due to their high central velocity peaks.  No ESF and MUGS lie well above the observations, but 120\% SN energy lies only slightly above the observations.

\subsection{Star Formation History}
\label{sec:sfh}
\begin{figure}
\resizebox{9cm}{!}{\includegraphics{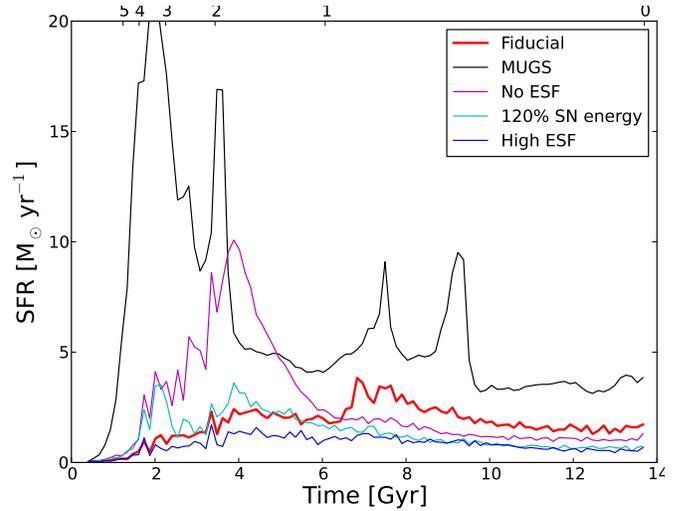}}
 \caption[Star Formation History]{ The star formation history of galaxies simulated with varying amounts of stellar feedback.  Here, the simulation starts at Time = 0 Gyr, on the left of this plot.  The star formation rate is drastically reduced with the increased stellar feedback.  The shape also changes from a quick rise and exponential falloff to a flatter evolution. }
\label{fig:sfh} 
\end{figure}

The star formation history in Figure \ref{fig:sfh} shows that varying feedback has a significant impact on when stars form.  The MUGS case shows a fairly standard star formation history for most simulations reported in the literature \citep{Scannapieco2012}.  Star formation follows the buildup of total halo mass as shown in Figure \ref{fig:4massev}.  All of the simulations without early stellar feedback experience a fairly broad peak of star formation just after 2 Gyr.  This is the time period when many nearly equal mass mergers are rapidly building the halo mass.  When the supernova energy is increased to 120\%, star formation after this peak is kept below 3 M$_\odot$ yr$^{-1}$.  Introducing the early stellar feedback reduces or eliminates this burst from the simulations shown here because the early stellar feedback prevents gas from collapsing into the central 2 kpc as shown in \S \ref{sec:center}.  Following the early peaks of star formation the star formation rate exponentially declines with intermittent bursts during mergers.  

The star formation rate has been calculated using 100 bins each 137 Myr wide in Figure \ref{fig:sfh}.  Figure \ref{fig:diffsfh} shows the star formation rate calculated using 25 Myr bins, which shows the bursty nature of star formation that is characteristic of all of the galaxies outside of the MUGS.  There are minor episodic bursts every 200-300 Myr overlaid onto the underlying shape of the star formation histories.  These minor bursts correspond to the dynamical time of the galaxy, approximately the time it takes for gas to collapse after it is heated by stellar feedback \citep[this phenomenon is described in more detail for dwarf galaxies in][]{Stinson2007}.

While the ESF galaxy does not show an increase in star formation at $z=4$ (2 Gyr), all the galaxies show a peak of star formation at $z=2$ (3.75 Gyr).  This corresponds to the time of the first merger of halos with masses $>10^{11}$ M$_\odot$.  The ``no ESF'' simulation (\emph{magenta line}) with $E_{SN}=10^{51}$ erg but without early stellar feedback undergoes a moderately long, 2 Gyr starburst due to this merger.  The fiducial simulation with early stellar feedback ($red$) has a shorter starburst ($\sim 100$ Myr) and smaller amplitude (8 M$_\odot$ yr$^{-1}$) at this time.  We note that a large population of small emission line galaxies has been detected at $z=2$ that have a major starburst as their first sign of significant star formation activity \citep{vanderWel2011}.

Shortly after the major merger, the fiducial simulation has another peak at 4 Gyr that corresponds to the accretion of gas remaining from the $z=2$ merger.  Its peak is lower and it lasts for a longer time.  There are a couple of small, gas rich mergers that create a second star formation rate peak at 7 Gyr.  This ends the merging history for g1536 and for the most part, the star formation rate slowly declines as star formation uses up the gas in the disk.  The simulations without early stellar feedback decline more rapidly than the fiducial simulation and have a lower star formation rate for the last 7 Gyr of the simulation.  This is due to those galaxies either using up their gas reservoirs faster (No ESF) or ejecting the gas further so that it is no longer available for star formation (120\% SN energy).

The star formation history of the fiducial simulation compares well to the star formation history of the Milky Way, though such studies have necessarily focused on the solar neighborhood \citep{RochaPinto1997,Rocha-Pinto2000,Hernandez2000,Cignoni2006}.  These studies find that the solar neighborhood has a mostly flat star formation rate with factors of a few variation and a peak at $z\sim\,1$.  Looking at nearby dwarf galaxies, \citet{Weisz2011} finds that star formation histories are also mostly flat with variations limited to factors of a few, which again compares well with the simulation.  We note that the galaxy in the \citet{Weisz2011} sample are all at least 3 magnitudes fainter than the simulated galaxies and have star formation rates two orders of magnitude lower.  So, in a general sense, the star formation history of the fiducial simulation agrees well with the observations.

\begin{figure}
\resizebox{9cm}{!}{\includegraphics{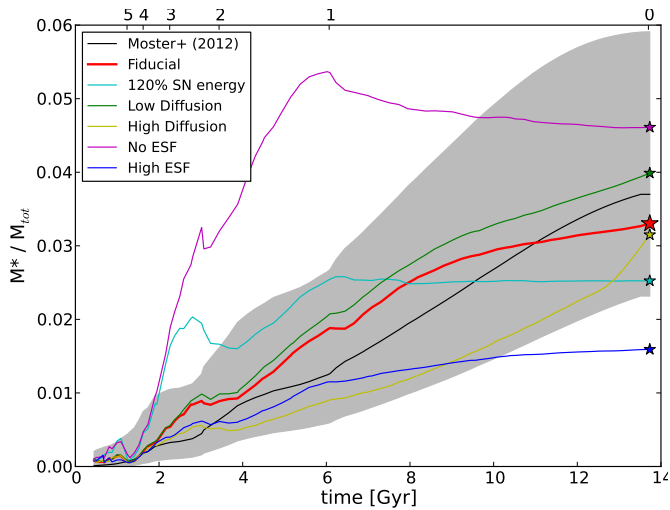}}
 \caption[Mstar Mhalo evolution]{ The evolution of the ratio of stellar mass to total halo mass for simulations using varying amounts of stellar feedback.  Here, the simulation starts at Time = 0 Gyr, on the left of this plot.  The black line is from \citet{Moster2012} that includes a detailed treatment of substructure to make the relationship between halo and stellar mass.  The grey region represents the $1\sigma$ variation in the model.  The simulations evolve mostly within the $1\sigma$ variation.  }
\label{fig:evguo} 
\end{figure}
Figure \ref{fig:evguo} presents another view of the evolution of stellar mass.  It shows the evolution of the ratio of stellar mass to total halo mass for simulations using varying amounts of stellar feedback.  The black line shows the evolution of the ratio based on the evolution of the total halo mass inside $r_{200}$ according the recently published stellar mass--halo mass relationship in \citet{Moster2012}.  \citet{Moster2012} adds a detailed treatment of substructure to match the abundances between dark matter halo and stellar mass as well as using updated high redshift luminosity functions.  The black line is calculated using the mass evolution of the g1536 halo including baryons.  The grey region represents the $1\sigma$ variation for the model.  The simulations that use early stellar feedback evolve mostly within the $1\sigma$ variation.  The fiducial ($red$) simulation follows the black line most closely throughout the evolution of the galaxy.  The simulations without early stellar feedback evolve outside the $1\sigma$ variation early in their evolution because of the large amount of early star formation.  A quick glance at Figure \ref{fig:sfh} indicates that the MUGS simulation that follows a star formation history typical of many published simulations would range even further outside the acceptable range of stellar masses at early times.  The 120\% SN energy simulation does not evolve very far outside the variance region, but it also does not follow the evolution of the mean line closely at all.  It is well above the mean for the first 7 Gyr and well below the mean for the last 5 Gyr.  It is possible that this unique evolution has to do with the uniquely quiet merger history of g1536, but it could also indicate a mismatch.  The high ESF model also evolves outside of the observed range of galaxy stellar masses. It forms fewer stars at $z=0$ than \citet{Moster2012} predict.

\subsection{Mass Evolution}
\label{sec:massev}
\begin{figure}
\resizebox{9cm}{!}{\includegraphics{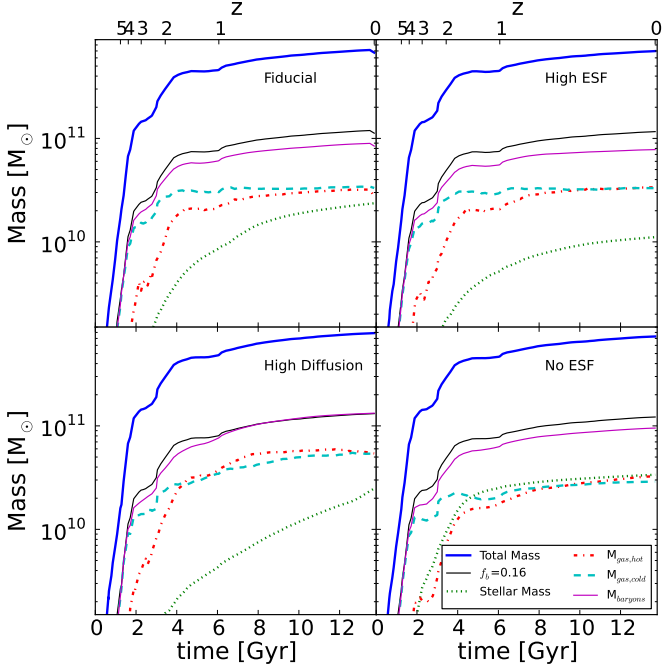}}
 \caption[Mass Evolution]{ The mass evolution of  the total halo mass ({\it blue}), stellar mass ({\it dotted-green}), cold gas (T$<10^5$ K, {\it dashed cyan}), and hot (T$> 10^5$ K, {\it dash-dotted red}) gas within $r_{vir}$ for the 4 simulations.  The thin black line shows the cosmic baryon fraction of the total halo mass, while the thin magenta line shows the actual baryonic mass content of the halo.}
\label{fig:4massev} 
\end{figure}

It is also possible to track the mass history for the other components of the galaxy.  Figure \ref{fig:4massev} shows the mass history for four of the simulations.  For each, the blue line represents the total galaxy mass inside $r_{vir}$, where $r_{vir}$ is the radius at which $\overline{\rho_{gal}} = 390 \rho_{background}$.  The black line is at one-sixth of the total mass, representing the cosmic fraction of baryons.  The magenta line shows the actual mass of baryons inside the galaxy.  The dashed cyan line represents the mass of gas with T$<10^5$ K while the dash-dotted red line represents the mass for T$>10^5$ K, and the dotted green line shows the evolution of stellar mass.  The total mass line clearly shows the two significant mergers at $z=2$ and $z=1$.  

The origin of hot gas mass is coincident with the onset of star formation, indicating that stellar feedback could play a role in creating the hot halo; it may not only be accretion shocks as described in \citet{Toft2002} and \citet{Crain2010}.  Proving this claim is beyond the scope of this paper, so we leave it for future work.  Both stellar mass and hot gas mass increase significantly during the $z=2$ merger, though the star formation increase is greatest in the case without early stellar feedback.  The no ESF simulation also evolves to  an equipartition between the three baryonic phases, stars, hot gas, and cold gas.  It appears that the other simulations are converging to this equipartition, but it is hard to predict whether the stars will reach the same value as hot and cold gas.  

In the two upper panels of simulations without thermal diffusion in particles with their cooling shut off, 25\% of the baryons escape the potential well of the halo during the $z=2$ merger.  Some baryons also initially escape in the simulations with thermal diffusion between all particles, but then baryons reaccrete.

\section{Discussion}
\label{sec:center}

\begin{figure}
\resizebox{9cm}{!}{\includegraphics{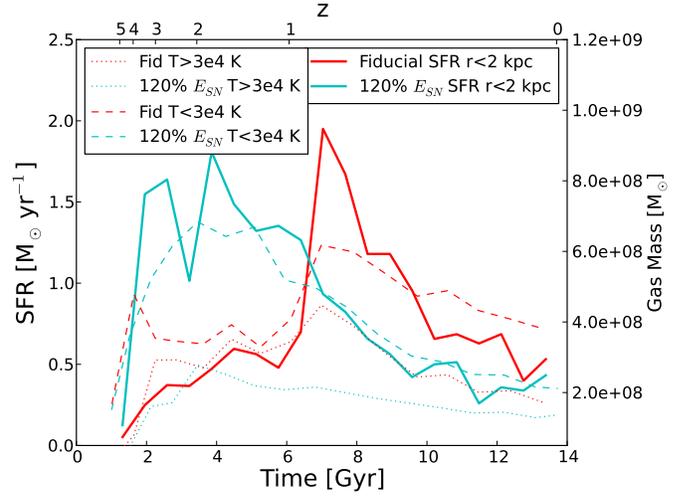}}
 \caption[Central Star Formation]{ The evolution of the inner 2 kpc of the stars and gas in the simulation for the fiducial (red) and 120\% SN Energy (cyan) models.  The solid lines show the star formation rate inside 2 kpc throughout the simulation.  The star formation rate scale is on the left y-axis.  The dashed lines show the cold (T$<3\times10^4$ K) gas mass evolution within 2 kpc.  The dotted lines show the hot (T$>3\times10^4$ K) gas mass evolution within 2 kpc.  The gas masses are given on the right y-axis.}
\label{fig:centsf} 
\end{figure}
The question remains about how the early stellar feedback more successfully reproduces observed galaxy properties than models without early stellar feedback.  The greatest difference seems to be in the galaxy centers; the simulations with early stellar feedback have flat rotation curves while those without have central peaks.  To examine this, Figure \ref{fig:centsf} compares the evolution of the inner 2 kpc of the fiducial simulation with the 120\% SN energy simulation.  In Figure \ref{fig:sfh}, it was shown that the 120\% SN energy simulation forms more stars early throughout the galaxy than the fiducial simulation.  Figure \ref{fig:centsf} shows that this pattern is replicated in the central 2 kpc, where most of the star formation is occurring.  Figure \ref{fig:centsf} also shows that the higher star formation follows the higher mass of cold gas in the central region of the 120\% SN energy simulation.  So, there is a clear difference in the cold gas and star formation inside the inner 2 kpc of the two simulations.

What is causing the difference in cold gas in the center?  There are several possiblities:
\begin{itemize}
 \item Cold gas is more efficiently ejected from the early stellar feedback simulation
 \item Cold gas is replaced by hot gas in the inner 2 kpc.
 \item The increased thermal pressure supports the gas at distances from the center of the galaxy further than 2 kpc.
\end{itemize}
Rather than the ESF simulation more efficiently ejecting gas, a look at the evolution of the galaxies\footnote{http://www.mpia-hd.mpg.de/$\sim$stinson/magicc/movies/c.1td.05rp.1/close.mp4 versus http://www.mpia-hd.mpg.de/$\sim$stinson/magicc/movies/esn1.2/close.mp4} shows that the 120\% SN Energy ejects gas at higher velocity and to larger radii than the early stellar feedback simulation.  The stronger gas ejection from the 120\% SN Energy simulation is also apparent in the star formation history.  The simulations without early stellar feedback have less star formation after $z\sim1$ than those with early stellar feedback because they have less gas remaining in the central region.  

Figure \ref{fig:centsf} also shows the mass of hot gas ($T>3\times10^4$ K, dotted lines) in the central 2 kpc as a function of time.  The hot gas mass is quite similar for the two simulations, so it is not $T>3\times10^4$ gas alone that is supporting gas from collapsing into the center.  

\begin{figure}
\resizebox{9cm}{!}{\includegraphics{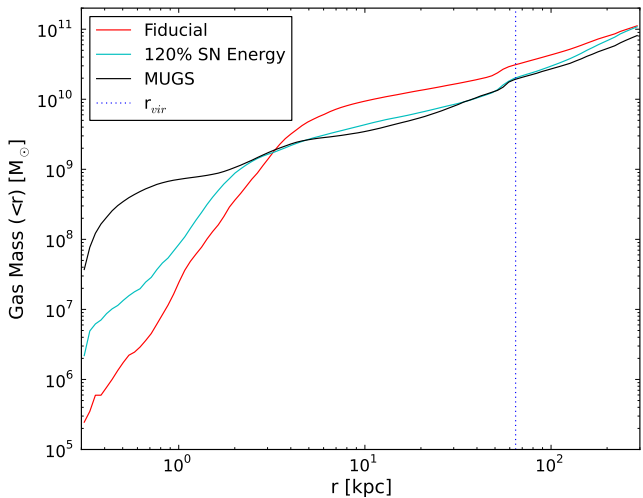}}
 \caption[z=2.25 Gas Cummulative Mass Profile]{The cumulative gas mass profile at $z=2.25$ for the fiducial simulation compared to the 120\% SN energy.  The 120\% SN energy simulation has up to ten times more mass at radii inside 3 kpc.  In the fiducial simulation, this mass is supported outside 3 kpc.}
\label{fig:gasprof} 
\end{figure}
Thus, we are left with the possibility that $T<3\times10^4$ K gas provides enough pressure to support gas to distances greater than 2 kpc.  To check this possibility, Figure \ref{fig:gasprof} shows the cumulative mass profile of gas at $z=2.25$ in the fiducial simulation compared with MUGS and the simulation that uses 120\% SN Energy.  $z=2.25$ is a time when there is a large discrepancy between the mass contained inside 2 kpc for the simulations with and without early stellar feedback.  There are also no satellites within 50 kpc of the disk at $z=2.25$.  Figure \ref{fig:gasprof} shows that there is less gas inside the central 2 kpc of the fiducial simulation than the 120\% SN Energy or MUGS simulations.  This rapidly changes outside 3 kpc as the early stellar feedback simulation has twice as much gas there.  The increase extends from 3 kpc well beyond $r_{vir}$.  In the case of the 120\% SN Energy simulation, the missing gas has been ejected out that far as the cumulative mass catches up to the fiducial simulation at 300 kpc.  In MUGS, however, the cumulative mass well below because the gas has been turned into stars.  

\begin{figure}
\resizebox{9cm}{!}{\includegraphics{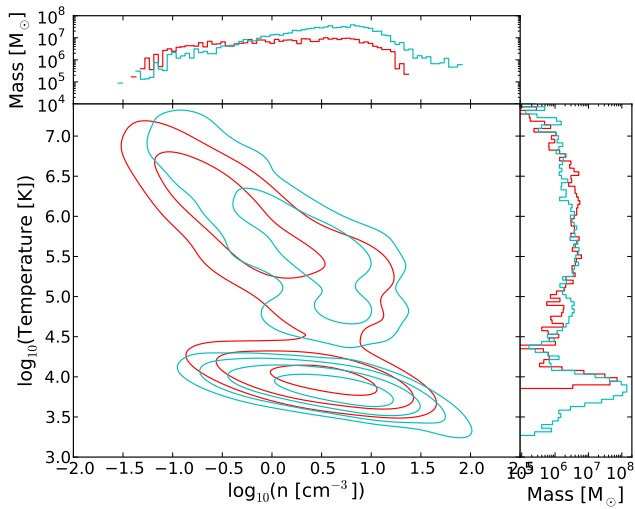}}
 \caption[Phase Diagram Comparison]{A comparison of the distribution of mass in temperature density phase space within 2 kpc of the galaxy centre.  The cyan contours represent the 120\% SN Energy simulation distribution.  The red contours show the fiducial simulation.  There are ten contour levels placed at logarithmic mass intervals.  Both the temperature axes have been collapsed into one-dimensional histograms along each axis.}
\label{fig:phasecompare} 
\end{figure}
To better understand why the fiducial simulation contains so much less gas inside 2 kpc than the 120\% SN Energy simulations, Figure \ref{fig:phasecompare} shows the temperature and density distributions for gas within 2 kpc in the two simulations.  Both the density and temperature distributions are projected into one-dimensional mass histograms in the top and right panels.  The temperature mass histogram shows that the peak of the gas mass distribution is at ISM conditions, $T<3\times10^4$ K, in both simulations.  It also shows that the fiducial simulation has little gas at $T<10^4$ K, whereas the 120\% SN Energy simulation has a significant amount.  This discrepancy shows how the early stellar feedback is having an effect similar to UV radiation by maintaining gas at $T>10^4$ K.  The UV heated gas at $10^4$ K will provide pressure support whereas the colder gas in the 120\% SN Energy simulation provides little.  Consequently, the 120\% SN Energy simulation contains more high density gas inside 2 kpc including gas up to 100 cm$^{-3}$.

The simulations have a similar amount of gas at $T>3\times10^4$ K, but the hot gas is lower density in the fiducial simulation.  This shows that the thermal feedback gas is released in lower density regions in the fiducial simulation and has an easier time adiabatically expanding. 

Our analysis shows the necessity for multiple forms of stellar feedback.  Adding SN Energy alone simply blows stronger winds, but does not create sufficient $10^4$ K gas to provide pressure support in the disk that keeps low angular momentum gas out of the center of the disk.  Without this feedback, the bulk of star formation happens too early.  Thus, it seems that winds alone are insufficient to prevent the formation of galaxies with high central concentrations.  It is also necessary for stellar feedback to create a warm component of the ISM.  That said, the fiducial simulation exhibits the same angular momentum distribution described in \citeauthor{Brook2012} (2012a), so outflows still play a key feedback role in shaping the galaxy.

\section{Conclusions}
We present a parameter study that varies the strength of stellar feedback in simulations of galaxy formation.  As a constraint, we use the stellar mass--halo mass relationship.  To reduce star formation enough to match the relationship, early stellar feedback from massive stars was required before they explode as supernova.  

Our fiducial model best fits the stellar mass--halo mass relationship, the evolution of that relationship, has a flat rotation curve and an exponential surface brightness profile with a modest bulge in the center.  In contrast to many previous cosmological galaxy formation simulations, most of the star formation occurs after $z=2$.  There are two peaks of star formation that correspond to the two significant mergers to the galaxy.  Both of these mergers are gas rich.  

We find that the mass of stars formed at $z=0$ is very sensitive to the amount and timing of stellar feedback employed.  Models that form too many stars follow a common pattern where they turn half of their baryons into stars, the other half is hot gas.  These galaxies also form many of their stars at higher redshift than what \citet{Moster2012} find when they compare high redshift luminosity functions with N-body simulations.  Galaxies that form too many stars have bright central concentrations of stars.  This is reflected in galaxy rotation curves as a high central peak.  Galaxies that form the right amount of stars have exponential surface brightness profiles and slowly rising rotation curves.

A simulation which does not use early stellar feedback, but increases the supernova energy to $1.2\times10^{51}$ erg ends up with a similar stellar mass to the fiducial simulation, but does not compare as well with other observed galaxy properties.  The difference is that our implementation of early stellar feedback keeps the gas above $10^4$ K and at densities below 10 cm$^{-3}$, an effect similar to that achieved by UV ionization.  The low density, warm gas keeps the disk extended prior to $z=1$ and thus keeps gas out of the central 2 kpc of the galaxy.  Since less gas makes it to the central region, fewer stars form before $z=1$ and, therefore, this model does not result in the massive central concentration that forms without early stellar feedback.

The early stellar feedback produces galaxies that correspond to observations in a number of ways.  Using early stellar feedback in dwarf galaxies spanning a wide mass range, \citeauthor{Brook2012a} (2012b) showed that the simulated galaxies match the disk scale length, gas fractions, and luminosities of observed galaxies.  \citeauthor{Brook2012} (2012a) has shown that the reason for such agreement is that outflows redistribute low angular momentum gas from the centers of galaxies to their outskirts.  \citet{Stinson2012} showed that the outflows that redistribute the gas also populate the circum-galactic medium with an amount of oxygen that closely matches observations of OVI in the CGM.  \citet{Maccio2012} showed that these outflows can also change the inner regions of dark matter density profiles from steeply rising cusps into flat cores.  In Kannan et al (\emph{in prep}), we will show how the stellar feedback described here performs in a larger sample of lower resolution galaxies in a cosmological volume.  Early stellar feedback helps keeps gas out of the center of galaxies, which leads to forming disk galaxies like those that we observe.

\section*{Acknowledgements}
We greatly appreciate the many suggestions the anonymous referee made to make this paper much better.
The analysis was performed using the pynbody package
(\texttt{http://code.google.com/p/pynbody}), which was written by Andrew Pontzen and Rok Ro\v{s}kar
in addition to the authors.  We thank Arjen van der Wel for useful conversations.

We are grateful to Ben Moster for providing his as yet unpublished stellar mass--halo mass relationship to us.  
The simulations were performed on the \textsc{theo} cluster of  the
Max-Planck-Institut f\"ur Astronomie at the Rechenzentrum in Garching;
the clusters hosted on \textsc{sharcnet}, part of ComputeCanada.  We greatly appreciate
the contributions of these computing allocations.
CBB acknowledges Max- Planck-Institut f\"ur Astronomie for its hospitality and financial support through the  Sonderforschungsbereich SFB 881 ``The Milky Way System''
(subproject A1) of the German Research Foundation (DFG).  AVM and GS also acknowledge support from SFB 881 (subproject A1) of the DFG.  JW and HMPC thank NSERC for support.  TQ acknowledges support from grant AST-0908499 from the NSF. 
\bibliographystyle{mn2e}
\bibliography{references}

\clearpage

\end{document}